\documentclass[amsmath,amssymb,prd,nofootinbib]{revtex4}

\usepackage{graphicx}
\usepackage{color}
\usepackage{subfigure}

\usepackage{hyperref}

\usepackage{citesort}
\include{aas_journals}

\usepackage{slashbox}

\newcommand{\be}{\begin{equation}}
\newcommand{\ee}{\end{equation}}
\newcommand{\bea}{\begin{eqnarray}}
\newcommand{\eea}{\end{eqnarray}}
\newcommand{\bean}{\begin{eqnarray*}}
\newcommand{\eean}{\end{eqnarray*}}

\newcommand{\Mpc}{\ensuremath{\,{\rm Mpc}}}

\newcommand{\eV}{\ensuremath{\,{\rm eV}}}

\begin{document}
\title{Cosmological parameter fittings with the BICEP2 data }
\author{Fengquan Wu$^1$, Yichao Li$^{1,2}$, Youjun Lu$^1$, Xuelei Chen$^{1,3}$}
\affiliation{$^1$National Astronomical Observatories, Chinese
Academy of Sciences, \\
20A Datun Road, Chaoyang District, Beijing 100012, China\\
$^2$ University of Chinese Academy of Sciences, 
Beijing 100049, China \\
$^3$ Center of High Energy Physics, Peking University, Beijing 100871, China
}

\date{\today}

\begin{abstract}
Combining the latest Planck, Wilkinson Microwave Anisotropy Probe (WMAP), 
and baryon acoustic oscillation (BAO) data,
we exploit the recent cosmic microwave background (CMB) 
B-mode power spectra data released 
by the BICEP2 collaboration to constrain the cosmological 
parameters of the $\Lambda$CDM model, esp. the primordial 
power spectra parameters of the scalar 
and the tensor modes, $n_s,\alpha_s, r, n_t$. 
We obtained constraints on the parameters for a lensed $\Lambda$CDM model using
the Markov Chain Monte Carlo (MCMC) technique, the marginalized 
$68\%$ bounds are
$ r   = 0.1043\pm_{  0.0914}^{  0.0307}, 
n_s = 0.9617\pm_{  0.0061}^{  0.0061},
\alpha_s =  -0.0175\pm_{  0.0097}^{  0.0105},
n_t  = 0.5198\pm_{  0.4579}^{  0.4515}.$ 
We found that a blue tilt for $n_t$ is favored slightly, but it
is still well consistent with flat or even red tilt. Our
$r$ value is slightly smaller than the one obtained by the BICEP team, 
as we allow $n_t$ as a free parameter without imposing the 
single-field slow roll inflation consistency relation. 
If we impose this relation, $r=0.2130\pm_{0.0609}^{0.0446}$.
For most other parameters,  the best fit values and measurement errors 
are not altered much by the introduction of the BICEP2 data.
\end{abstract}

\pacs{}

\keywords{BICEP2,B-mode, inflation}

\maketitle


\section{Introduction}
The BICEP2 experiment\cite{Ade:2014xna,Ade:2014gua}, 
dedicated to the observation of the 
cosmic microwave background (CMB) polarization,
has announced the detection of the primordial B-mode polarization, 
from observations of about 380 square degrees 
low-foreground area of sky during 2010 to 2012 in the South Pole.
The detected B-mode power is in the multipole range $30<\ell<150$, 
a clear excess over the base lensed-$\Lambda$CDM  model at these 
small $\ell$s, this excess can not be explained by the lensing 
contribution, for the CMB lensing contribution to B-mode polarization 
peaks at  $\ell\sim1000$, while the contributed power at $\ell\sim100$,
is small. The BICEP team has  also examined possible systematic error and 
potential foreground contaminations and excluded these as possible 
source of the observed B-mode power. The 
cross-correlation between frequency bands shows little
change in the observed amplitude, implying that frequency-dependent 
foreground are not the dominant contributor.  
The presence of the B-modes induced by the primordial gravitational 
wave in the early universe provides a direct evidence for the inflation theory.

The tensor mode contribution to the CMB anisotropy may affect the 
global fitting of the cosmological parameters. The BICEP group reported their
measured value of tensor-to-scalar ratio as
 $r=0.20^{+0.07}_{-0.05}$, based on the lensed-$\Lambda$CDM+tensor 
model, and derived from importance sampling of the Planck MCMC chain 
using the direct likelihood method, but they did not give constraints
on other parameters. The unexpectedly large tensor-to-scalar 
ratio inspires a lot of interests on re-examining the
inflation models
\cite{ Hertzberg:2014aha,Choudhury:2014kma,Ma:2014vua,Gong:2014cqa,Xia:2014tda,Cai:2014bda}
and observation datasets\cite{Zhao:2014rna,Zhao:2010ic,Zhang:2014dxk}.

In this paper, we use the newly published BICEP2 CMB B-mode data,
combined with the Planck CMB temperature data\cite{Collaboration:2013uv},
the WMAP 9 year CMB polarization data\cite{Hinshaw:2013dd, Bennett:2013ew},
and the BAO data from the SDSS DR9\cite{Anderson:2013jb}, 
SDSS DR7\cite{Padmanabhan:2012ft}, 6dF\cite{Beutler:2011ea}, 
to constrain the cosmological parameters in the lensed $\Lambda$CDM model. 
We derive constraints on the lensed $\Lambda$CDM  model 
using the publicly available code COSMOMC\cite{Lewis:2002ah},
which implements a Metropolis-Hastings algorithm to perform 
a MCMC simulation in order to fit the cosmological
parameters. This method also provides reliable error estimates
on the measured variables.

Previous CMB observations from the Planck satellite, the WMAP satellite and 
other CMB experiments yielded a limit of much smaller 
tensor-to-scalar ratio $r<0.11$ (at $95\%$ C.L.)\cite{Collaboration:2013uv},
so there is some tension between these and the BICEP result
at least in the simplest lensed $\Lambda$CDM+tensors model. 
As pointed out by the BICEP team\cite{Ade:2014xna}, a simple way to 
relax this tension is to take the running of spectral index into 
account, we will explore this possibility in our fit. 
There are also wide spread 
interests in the tensor power spectral index, as it 
is an important additional source of information for 
distinguishing inflation models \cite{2014arXiv1403.5922A,2014arXiv1403.5163G,2014arXiv1403.4927B}, and a blue tensor power spectrum tilt $n_t\sim 2$ have
been reported using the B-mode measurement\cite{2014arXiv1403.5732G}. Here
we shall also investigate this problem and obtain an estimate of 
$n_t$ and its measurement error.

\section{The fitting of cosmological parameters}
We explore the cosmological parameter space and
obtain limits on cosmological parameters by using the MCMC technique with 
the CosmoMC code \cite{Lewis:2002ah}. In our simulation we 
collected about 500000 chain samples, the first 1/3 of the data 
was used for burning in the chains and not used for the final analysis. 
In addition of the BICEP data \cite{Ade:2014xna}, we use the
 Planck CMB temperature data\cite{Collaboration:2013uv},
the WMAP 9 year CMB polarization data\cite{Hinshaw:2013dd, Bennett:2013ew},
and the BAO data from the SDSS DR9\cite{Anderson:2013jb}, 
SDSS DR7\cite{Padmanabhan:2012ft}, 6dF\cite{Beutler:2011ea} in our cosmological
parameter fitting. Below, we use the following labels to denote the 
different data sets included in the fitting:
\begin{itemize}
  \item Planck + WP : Planck high $\ell$, low $\ell$\cite{Collaboration:2013uv}, and WMAP9 polarization data\cite{Hinshaw:2013dd, Bennett:2013ew}.
 \item Planck + WP + BAO :   add BAO data from SDSS DR9\cite{Anderson:2013jb},  
 SDSS DR7\cite{Padmanabhan:2012ft}, 6dF\cite{Beutler:2011ea}. 
 \item Planck + WP +BAO + BICEP :  add BICEP data\cite{Ade:2014xna,Ade:2014gua}.  
\end{itemize}

As noted by the BICEP group, in order to be compatible with the 
Planck data, the running of spectrum tilt $\alpha_s$ is needed. 
We shall consider a $\Lambda$CDM model, and 
assume that the scalar perturbations are purely adiabatic, and the scalar and
tensor mode power spectra are parameterized by 
\begin{eqnarray}
  P_\zeta \left( k \right) &\equiv& A_s \left( \frac{k}{k_0} \right)^{n_s - 1 + \frac{1}{2} \alpha_s \ln \frac{k}{k_0} } \;\; \ , 
  \label{parametrizationP}\\
  P_t \left( k \right) &\equiv& A_t \left( \frac{k}{k_0} \right)^{n_t  } \;\; \ ,
 \label{parametrizationPT}
\end{eqnarray}
where $k_0 = 0.05$ Mpc$^{-1}$, is the pivot scale, 
roughly in the middle of the logarithmic 
range of scales probed by the WMAP and Planck experiments.   The parameter
$\alpha_s$ denotes the running of the scalar 
spectral tilt\cite{1995PhRvD..52.1739K} 
with $\alpha_s =  d \, n_s / d \, {\rm ln }  \, k$.
The primordial tensor-to-scalar ratio is defined by $r\equiv A_t /A_s$ at a
chosen pivot scale, for example $r_{0.05}$ is defined 
at $k_0=0.05 \Mpc^{-1}$, and $r_{0.002}$ at $k_0=0.002 \Mpc^{-1}$. 
The relation between $r_{0.05}$ and $r_{0.002}$ could be inferred from Eq.\ref{parametrizationP} and Eq.\ref{parametrizationPT}:
\begin{eqnarray}
  r_{0.002}=r_{0.05} \frac{0.04^{n_t  } }{ 0.04^{n_s - 1 + \frac{1}{2} \alpha_s \ln 0.04 } }  ~  .
  \label{eq:r0.05-r0.002}
\end{eqnarray}
Throughout 
this paper, $r$ without the subscript (as in our plots) 
is $r_{0.002}$. In the Planck Collaboration paper 
XVI (2013)\cite{Collaboration:2013uv}, $n_t$ is 
assumed to be close to zero, and satisfies a single field slow roll inflation 
consistency relation 
\begin{equation}
n_t = -\frac{r}{8} ~ .
\label{eq:consistency_relation}
\end{equation} 
Note that in Eq.(\ref{eq:consistency_relation}), $n_t$ and $r$ should be 
defined at the same pivot scale.
The BICEP group adopted the same assumption, and applied the importance
sampling method on the Planck MCMC chains \cite{Collaboration:2013uv} with the 
addition of the B-mode data to obtain constraints on $r$ and $n_s$ \cite{Ade:2014xna}. 
Here we study the more general case, with $n_t$ and $r$ treated 
as independent parameters.
We fix the the number of neutrinos as $N_{eff}=3.046$, and the 
sum of neutrino masses as the Planck best fit $\sum m_\nu =0.06 \eV$.
The lensing amplitude parameter $A_L$ is fixed to $1$,
and we put flat priors on all fitting  parameters. 

\begin{figure}[tbp]
\begin{center}
\includegraphics[width=0.6\textwidth]{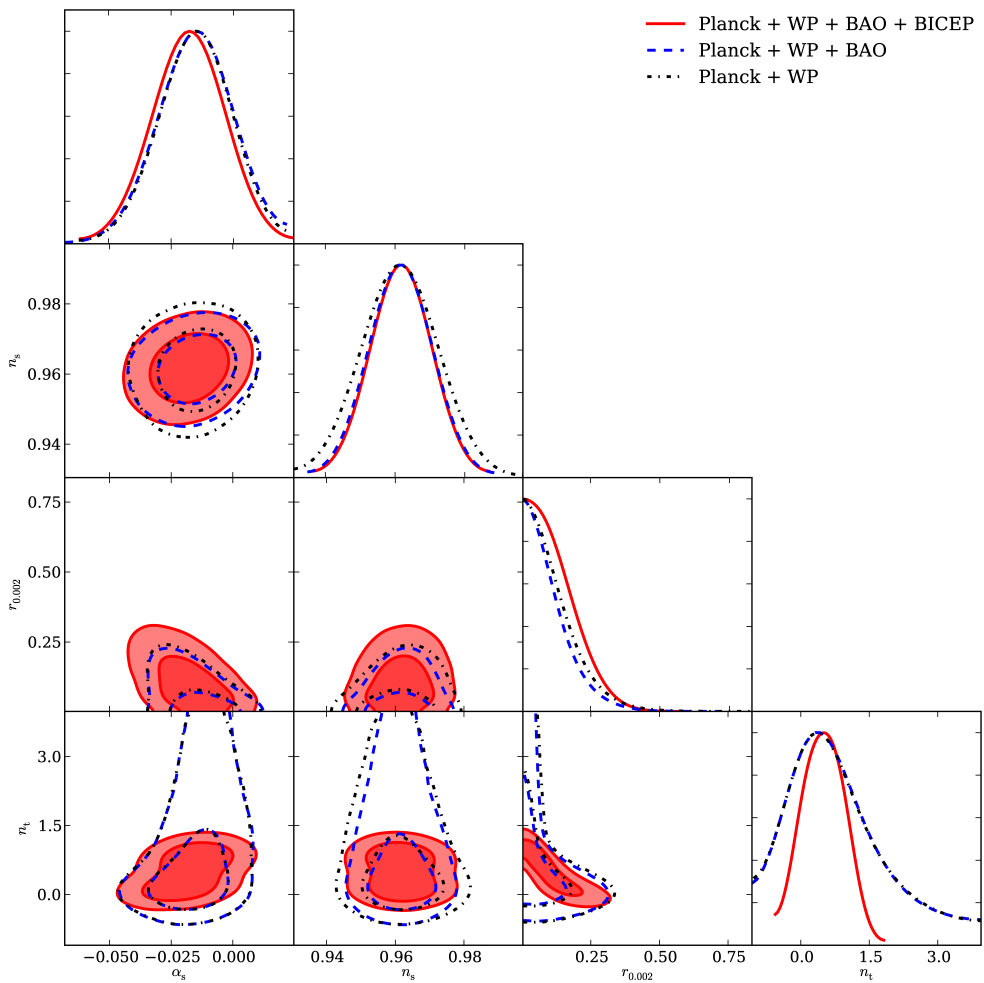}
\caption{Joint constraints on primordial power spectrum parameters.}
\label{fig:triangle-pw}
\end{center}
\end{figure}
\begin{figure}[htbp]
\begin{center}
  \includegraphics[width=0.6\textwidth]{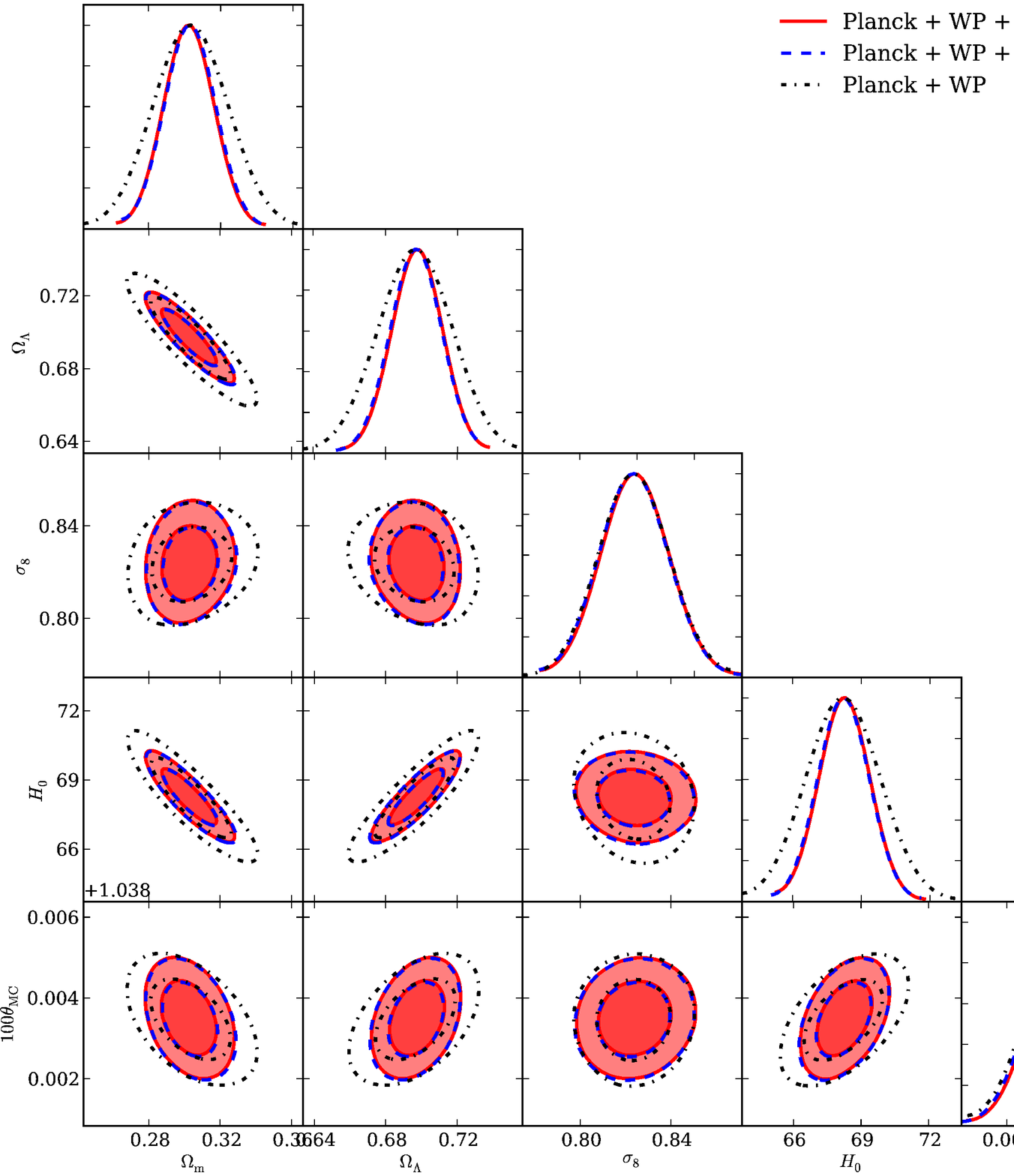}
\caption{Constraints on other cosmological parameters}
\label{fig:triange-other-param}
\end{center}
\end{figure}

\begin{figure}[htbp]
\begin{center}
\includegraphics[width=0.6\textwidth]{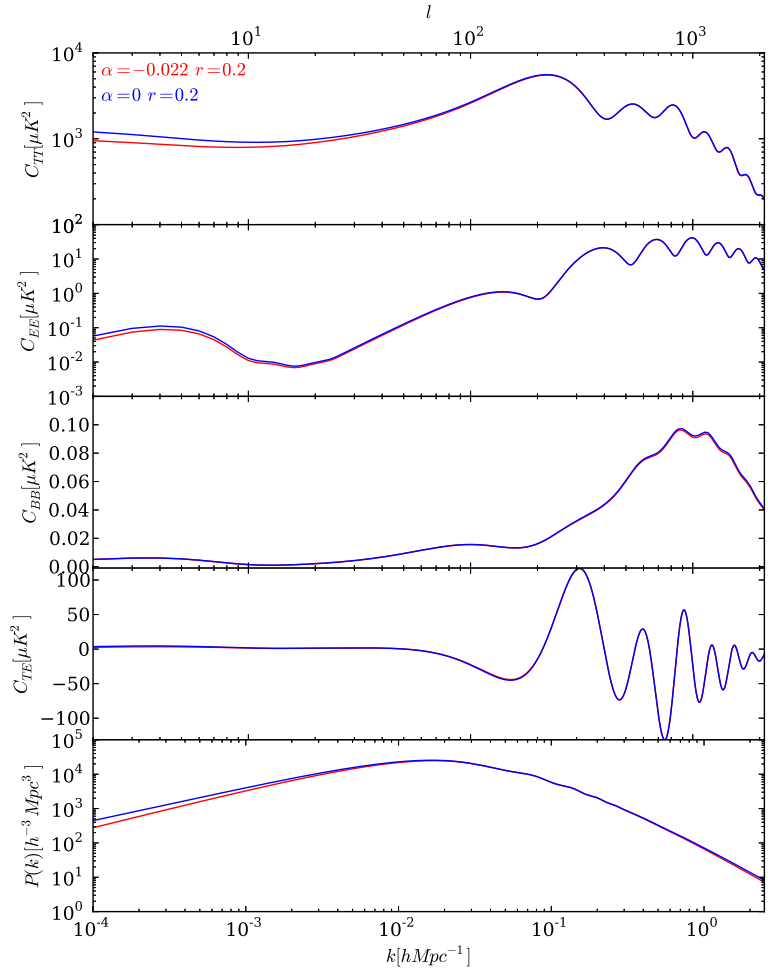}
\caption{Comparison of CMB and matter power spectra for 
different running index $\alpha$.}
\label{fig:diff_alpha}
\end{center}
\end{figure}

The constraints on the primordial power spectrum parameters are shown 
 in Fig.\ref{fig:triangle-pw}, 
and constraints on other parameters in Fig.\ref{fig:triange-other-param}.  
The marginalized $68\%$ bounds on the parameters based on different datasets 
are listed in Table \ref{tb:params}.
For the scalar spectral index and running,  using the Planck + WP  + BAO + BICEP dataset, 
we obtain 
$$n_s=0.9617 \pm {0.0061}, \qquad \alpha_s =  -0.0175\pm_{  0.0097}^{  0.0105}$$
while for the Planck + WP + BAO data 
$$n_s=0.9616 \pm{0.0061}, \qquad \alpha_s = -0.0148\pm_{  0.0095}^{  0.0108}$$ 
(see Table \ref{tb:params}),  the best-fit value of $\alpha_s$ 
is $0.027$ smaller after including the BICEP2 data.
The decrease in $\alpha_s$ reduces the $TT$ angular power at 
small $\ell$ region (see Fig.\ref{fig:diff_alpha}), and helps to alleviate
the tension between the high $r$ value obtained with BICEP data and the limit derived
from the large scale $TT$ auto-correlation power from Planck experiment.    
The  effect of decreasing $\alpha_s$ on the other power spectra is that it  
could  lower the matter power spectrum at very large scale 
and very small scales. If future galaxy surveys 
can probe the matter power spectrum at extremely large scales,
the constraint on $\alpha_s$ can be further improved.

\begin{figure}[htp]
\begin{center}
\includegraphics[width=0.35\textwidth]{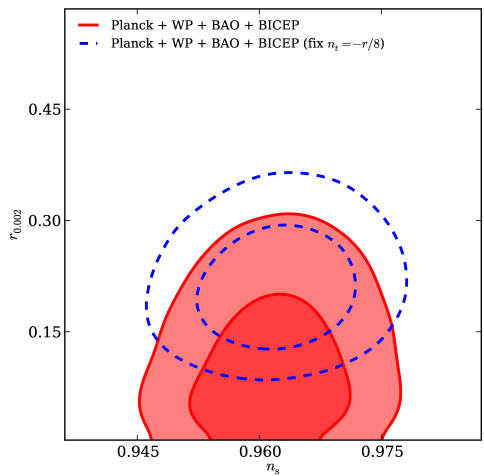}
\caption{Comparison of  the two dimensional likelihood contour obtained by 
using and  without using single-field inflation consistency relation. 
The dataset used is Planck + WP  + BAO + BICEP.}
\label{fig:contour_r02_ns}
\end{center}
\end{figure}

For the tensor-to-scalar ratio $r$, the BICEP group 
reported a value of $r = 0.20^{+0.07}_{-0.05}$ based on their fit
to the CMB power spectrum, with the consistency relation 
Eq.(\ref{eq:consistency_relation}).  
Using the Planck + WP  + BAO + BICEP dataset, we obtain
$$ r   = 0.1043\pm_{  0.0914}^{  0.0307},$$
where we have taken $n_t$ and $r$ as independent free parameters.  
If we fix $n_t$ to the single-field inflation consistency relation value
as the Planck and BICEP2 group did,
we will obtain a higher value of $r$ and can also
place a tighter constraint on $r$.
We compare the two dimensional likelihood contour of $r$ vs $n_t$ 
with and without  single-field inflation consistency relation 
in Fig.\ref{fig:contour_r02_ns}. 
Blue dashed curves in  Fig.\ref{fig:contour_r02_ns} show the result 
if we impose the single-field slow roll inflation consistency relation 
$n_t=-r/8$ in the fitting, with 
$r=0.2130\pm_{0.0609}^{0.0446}$(1$\sigma$ error bar). Note that with the
consistency relation, the $r$ value is significantly higher than the
one without. 

\begin{figure}[htb]
\begin{center}
\includegraphics[width=3in]{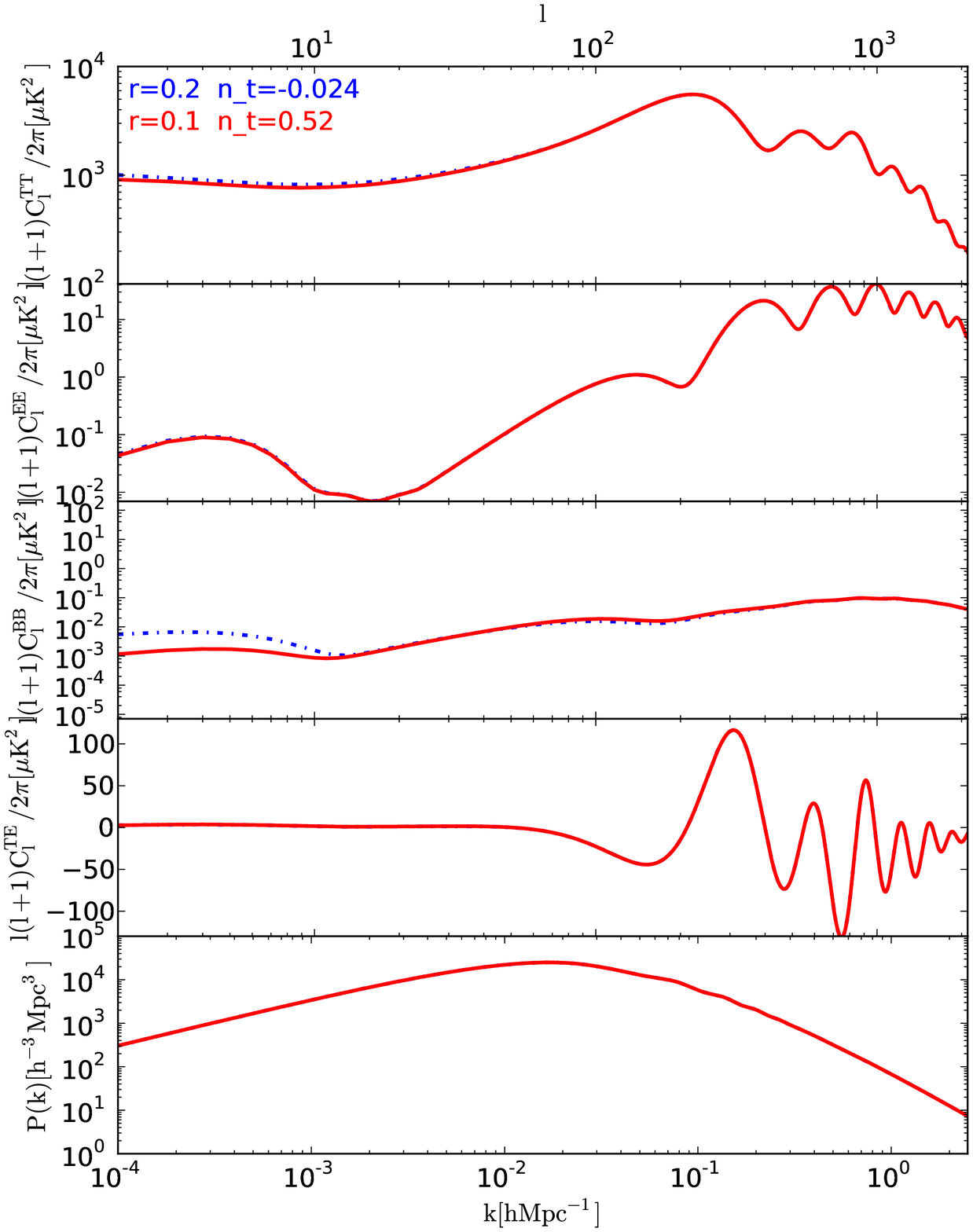}
\caption{Comparison of power spectra results predicted by two $n_t$ and $r$ parameter sets.
One set is our best fitting value(red curves), another is from BICEP2 paper\cite{Ade:2014xna}(blue dotted curve). }
\label{fig:diff_nt_r}
\end{center}
\end{figure}

For the tensor spectrum index  $n_t$, using the Planck + WP  + BAO + BICEP 
dataset and without consistency relation, we obtain
$$ n_t  = 0.5198\pm_{  0.4579}^{  0.4515}.$$
This result shows that a blue tilt is slightly favored,
but a flat or even red tilt is still consistent with the current data.

So, without imposing the consistency relation, 
the best fitting value of the $r_{0.002}$ and $n_t$ we obtain are $r \sim 0.1$ 
and $n_t \sim 0.52$, while the BICEP group reported
$r \sim 0.2$ and $n_t \sim -0.024$ (fixed by simple-field slow-roll 
inflation consistency relation prior) \cite{Ade:2014xna}. 
We plot the CMB power spectra and matter power spectra according 
to these two fits in Fig.\ref{fig:diff_nt_r}.
The red curves and blue curves overlap each other for most $\ell$-values, the 
main difference is at the very large scales($\ell<15$), where the statistics
are poor, and it is hard to distinguish the two cases with the current
the current observational data.

Constraints on the other cosmological parameters are shown 
in Fig.\ref{fig:triange-other-param} and Table \ref{tb:params}.
By combining the BAO data, we obtain tighter constraints on all parameters,
and help break the parameter degeneracy. Nevertheless, there is not 
much difference for these parameters after the addition of the BICEP data.

\begin{table*}
    \begin{center}
    \caption{\label{tb:params}
        Summary of the best fit values of cosmological parameters and the corresponding
        68\% intervals. The ``Best fit'' column is the best fitting value inferred from minimum $\chi^2$ point in the whole MCMC chains.
        The center value in ``$68\%$ limits'' column is the parameter according to  the peak in one-dimensional 
        likelihood distribution.
        The ``Planck + WP" column lists the result of using the 
        temperature map from Planck and polarization map form WMAP9; 
        ``Planck + WP + BAO" column shows the result with BAO data added; 
        ``Planck + WP + BAO+ BICEP" column is the result with BICEP2 
data added. 
    }
\begin{tabular}{c|cc|cc|cc} \hline\hline
& \multicolumn{2}{|c}{Planck + WP + BAO + BICEP}& \multicolumn{2}{|c}{Planck + WP + BAO}& \multicolumn{2}{|c}{Planck + WP}\\
\hline
Parameter & Best fit & $68\%$ limits& Best fit & $68\%$ limits& Best fit & $68\%$ limits\\
\hline
$n_\mathrm{s}$  & $  0.9618$ & $  0.9617\pm_{  0.0061}^{  0.0061}$ & $  0.9591$ & $  0.9616\pm_{  0.0061}^{  0.0061}$ & $  0.9631$ & $  0.9614\pm_{  0.0073}^{  0.0072}$ \\
$r_{0.002}$  & $  0.0634$ & $  0.1043\pm_{  0.0914}^{  0.0307}$ & $  0.0015$ & $<  0.0649$ & $  0.0117$ & $<  0.0684$ \\
$\alpha_\mathrm{s}$  & $ -0.0080$ & $ -0.0175\pm_{  0.0097}^{  0.0105}$ & $ -0.0129$ & $ -0.0148\pm_{  0.0095}^{  0.0108}$ & $ -0.0090$ & $ -0.0150\pm_{  0.0094}^{  0.0109}$ \\
$n_\mathrm{t}$  & $  0.7293$ & $  0.5198\pm_{  0.4579}^{  0.4515}$ & $  0.6230$ & $  0.8324\pm_{  1.1676}^{  0.3823}$ & $  1.3546$ & $  0.8361\pm_{  1.1803}^{  0.3868}$ \\
$\Omega_{\mathrm{m}}$  & $  0.3015$ & $  0.3026\pm_{  0.0094}^{  0.0094}$ & $  0.3080$ & $  0.3032\pm_{  0.0095}^{  0.0095}$ & $  0.3095$ & $  0.3037\pm_{  0.0138}^{  0.0137}$ \\
$\Omega_\Lambda$  & $  0.6985$ & $  0.6974\pm_{  0.0094}^{  0.0094}$ & $  0.6920$ & $  0.6968\pm_{  0.0095}^{  0.0095}$ & $  0.6905$ & $  0.6963\pm_{  0.0137}^{  0.0138}$ \\
$\sigma_8$  & $  0.8173$ & $  0.8242\pm_{  0.0099}^{  0.0099}$ & $  0.8218$ & $  0.8237\pm_{  0.0098}^{  0.0098}$ & $  0.8303$ & $  0.8236\pm_{  0.0099}^{  0.0099}$ \\
$H_0$  & $ 68.2990$ & $ 68.2599\pm_{  0.7416}^{  0.7430}$ & $ 67.8668$ & $ 68.2224\pm_{  0.7437}^{  0.7472}$ & $ 67.6910$ & $ 68.2043\pm_{  1.0658}^{  1.0663}$ \\
$100\theta_{\mathrm{MC}}$  & $  1.0417$ & $  1.0415\pm_{  0.0006}^{  0.0006}$ & $  1.0413$ & $  1.0415\pm_{  0.0006}^{  0.0006}$ & $  1.0411$ & $  1.0415\pm_{  0.0006}^{  0.0006}$ \\
\hline\hline
\end{tabular}
\end{center}
\end{table*}

\section{Conclusion}

In this paper, we use the newly published BICEP2 CMB B-mode data,
 Planck CMB temperature data\cite{Collaboration:2013uv},
 the WMAP 9 year CMB polarization data\cite{Hinshaw:2013dd, Bennett:2013ew}
to constrain the base lensed $\Lambda$CDM model. In addition to the CMB data, 
we also use the BAO data  
from SDSS DR9\cite{Anderson:2013jb}, SDSS DR7\cite{Padmanabhan:2012ft}, 
6dF\cite{Beutler:2011ea}, which help to break parameter degeneracy.

For most parameters, we find that the best fit values and 
measurement errors are not altered much by the introduction of 
the BICEP2 data. The most affected parameters are $r$, $\alpha_s$ and $n_t$. 
Combining Planck + WP  + BICEP+ BAO  dataset, we obtain marginalized
$68\%$ bounds on  some interested parameters are:
\begin{eqnarray}
 r  &=& 0.1043\pm_{  0.0914}^{  0.0307} ~ , \\
 n_s &=& 0.9617\pm_{  0.0061}^{  0.0061} ~  , \\
 \alpha_s &=&  -0.0175\pm_{  0.0097}^{  0.0105} ~ , \\
    n_t  &=& 0.5198\pm_{  0.4579}^{  0.4515} ~ .
  \label{eq:final}
\end{eqnarray}

We find that a blue tensor tilt ($n_t>0$) is slightly favored, 
but a flat or red tilt is consistent with the data.  
The best fitting value of $r$ we obtain is slightly smaller 
than BICEP2 group obtained,and the constraint on $r$ is also 
looser than BICEP2 group obtained. This result is reasonable, 
as we have not imposed the single-field-slow-roll inflation 
consistency relation on $n_t$, and treated it as an independent 
parameter. If we impose this relation, we will
obtain $r=0.2130\pm_{0.0609}^{0.0446}$($1\sigma$ error) instead.

In the near future, Planck and other experiments will
provide more data on CMB polarization, and help improve the constraint 
on these parameters.

\section*{Acknowledgements}

We thank  Antony Lewis for kindly providing us the beta version 
of  CosmoMC code for testing.
Our MCMC computation was performed on the Laohu cluster in 
NAOC and on the GPC supercomputer at the SciNet HPC Consortium.
This work is supported by the Chinese Academy 
of Science Strategic Priority Research Program ``The Emergence of
Cosmological Structures'' Grant No. XDB09000000,  by  the 
NSFC grant 11103027, 11373030 and
the Ministry of Science and Technology 863 project grant 2012AA121701.

\bibliography{bicep2.bib,wufq-cmb-b-mode.bib}
\bibliographystyle{utphys}

\end{document}